\documentclass[12pt,a4 paper]{article} 
\usepackage{amsfonts,amsmath,amsthm,amscd,amssymb,latexsym}
\usepackage{graphics}
\usepackage{tikz}
\usetikzlibrary{matrix}
\usepackage{epsfig}

\usepackage{amsmath,amsthm,amssymb}
\newtheorem{theorem}{Theorem}[section]

\newtheorem{example}[theorem]{Example}
\theoremstyle{definition}

\theoremstyle{remark}

\numberwithin{equation}{section}

\begin{document} 
\title{\textbf{On the $\mathbb Z_{q}$-MacDonald Code and its Weight Distribution of dimension $3$}}


\maketitle

\author{\textbf{J. Mahalakshmi}$^{1}$ and \textbf{C. Durairajan}$^{2}$\\\\
\footnotesize $^{1,2}${Department of Mathematics, Bharathidasan University, Tiruchirappalli, Tamil Nadu \\Pin - 620024, India.\\
\footnotesize {\textbf{$^{1}$mathmaha1984@gmail.com} and
\footnotesize {\textbf{$^{2}$cdurai66@rediffmail.com.}}\\\\\\

\noindent{\bf Abstract:} \  In this paper, we determine the parameters of $\mathbb Z_{q}$-MacDonald Code of dimension $k$ for any positive integer $q\geq 2$. 
Further, we have obtained the weight distribution of $\mathbb Z_{q}$-MacDonald code of dimension 3 and furthermore, 
we have given the weight distribution of $\mathbb {Z}_q$-Simplex code of dimension $3$ for any positive integer $q\geq 2$. \\


\section{Introduction}
A code $C$ is a subset of $\mathbb Z_{q}^{n}$ where $\mathbb Z_{q}$ is the set of all integers modulo $q$ and $n$ is any positive integer. 
Let $x, y\in\mathbb{Z}_{q}^{n}.$ Then the {\it Hamming distance} between $x$ and $y$ is the number of coordinates in which they differ. It is denoted by $d(x,y).$
Vividly $d(x,y)=wt(x-y),$ the number of non-zero coordinates in $x-y$ is called the {\it Hamming weight} of $x-y.$ The {\it minimum Hamming distance} $d$ of $C$ is
defined as $$d=\min\{d(x,y) \mid x,y\in C \ and\ x\neq y \}=\min\{wt(x-y) \mid x,y\in C \ and\ x\neq y \}$$ 
and the {\it minimum Hamming weight} of C is $\min\{wt(c) \mid c \in C \text{ and } c\ne 0\}.$ 
Hereafter we simply call the minimum Hamming distance and the minimum Hamming weight, the minimum distance and the minimum weight respectively.
A code over $\mathbb Z_{q}$ of length $n$, cardinality $M$ with the minimum distance $d$ is called an $(n, ~M, ~d)$
$\mathbb{Z}_q$-code. Let $C$ be an $(n, ~M, ~d)$ $\mathbb {Z}_q$-code. For $0\leq i \leq n,$ let $A_{i}$ be the number of codewords of the Hamming weight $i.$ Then $\{A_{i}\}_{i=0}^{n}$ is called the {\it weight distribution} of the code $C.$

We know pretty well that $\mathbb Z_{q}$ is a group under the addition modulo $q$. Then $\mathbb Z_{q}^{n}$ is a
group under coordinate-wise addition modulo $q$. $C$ is said to be a {\it $\mathbb Z_{q}$-linear code }if $C$ is a subgroup of $\mathbb Z_{q}^{n}.$  
In fact, it is a free $\mathbb Z_{q}$-module. Since $\mathbb Z_{q}^{n}$ is a free $\mathbb Z_q$-module, it has a basis. Therefore, every
$\mathbb Z_q$-linear code has a basis. Since $\mathbb Z_{q}^{n}$ has a finite basis, $\mathbb Z_{q}$-linear code has a finite dimension.
Since $\mathbb Z_{q}^{n}$ is finitely generated $\mathbb Z_{q}$-module, it implies that $C$ is a finitely generated submodule of $\mathbb Z_{q}^{n}.$ 
The cardinality of a minimal generating set of $C$ is called the rank of the code $C$ \cite{r9}. A generator matrix of $C$ is a matrix the rows of which generate $C$.
Any linear code $C$ over $\mathbb Z_{q}$ with generator matrix $G$ is permutation-equivalent to a code with generator matrix of the form
\begin{equation*}
\left[\begin{array}{ccccccc}
I_{k} & A_{01} & A_{02} & \cdots & A_{0s-1} & A_{0s} \\  
0 & z_{1}I_{k_{1}} & z_{1}A_{12} & \cdots & z_{1}A_{1s-1} & z_{1}A_{1s} \\
0 & 0 & z_{2}I_{k_{2}} & \cdots & z_{2}A_{2s-1} & z_{2}A_{2s} \\
\vdots & \vdots & \vdots & \ddots & \vdots & \vdots \\
0 & 0 & 0 & \cdots & z_{s-1}I_{k_{s-1}} & z_{s-1}A_{s-1s}
\end{array}\right]
\end{equation*}
where $A_{ij}$ are matrices over $\mathbb Z_{q}$, $\{z_{1}, ~z_{2}, ~\cdots, ~z_{s-1}\}$ are the zero-divisors in $\mathbb Z_{q}$ and the columns are grouped into blocks of sizes $k, ~k_{1}, ~\cdots, ~k_{s-1}$ respectively. 
Then $|C|=q^{k}(\frac{q}{z_1})^{k_{1}}(\frac{q}{z_2})^{k_{2}}\cdots(\frac{q}{z_{s-1}})^{k_{s-1}}$. If $k_{1}=k_{2}=\cdots=k_{s-1}=0,$ then the code $C$ is called $k$-dimensional code. Every $k$ dimension $\mathbb Z_{q}$-linear code with length $n$ and the minimum distance $d$ is called an 
{\it $[n, ~k, ~d]$ $\mathbb Z_q$-linear code.}

There are many researchers doing research on codes over finite rings \cite{r1}, \cite{byi}, \cite{dk10}, \cite{eby} and \cite{vsr96}. In the last decade, there have been many number of researchers doing research on codes over $\mathbb Z_{4}$ and $\mathbb Z_{q}$  
\cite{bgl99}, \cite{cs93}, \cite{r12}, \cite{r13} and \cite{r8}. Further, in \cite{r6}, they have determined the parameters of $\mathbb Z_q$-Simplex codes of dimension $k$ and in \cite{r7}, they have obtained the weight distribution of $\mathbb Z_q$-Simplex codes of dimension $2$ for any positive integer $q\geq 2.$
\begin{theorem}\cite{r6}
 The $\mathbb Z_{q}$-Simplex code of dimension $k$ is an $[n_{k}=\frac{q^{k}-1}{q-1}, ~k, ~d_{k}=\frac{q}{p}(p-1)n_{k-1}+1]$ $\mathbb Z_q$-linear code where $p>1$ is the smallest divisor of $q$.
\end{theorem}

In \cite{cpcd}, they have defined a  $\mathbb Z_{q}$-linear code which is similar to the MacDonald 
code over finite field. But it gives different weight distribution. In the generator matrix
$G_{k}(q)$ of  $\mathbb Z_{q}$-Simplex code $S_{k}(q)$ of dimension $k,$ by deleting the
matrix 
\[ 
 \left[\begin{array}{c}
  \Large{O}\\
  G_u(q)
 \end{array}
\right]\] where $2 \leq u \leq k - 1$ and $\Large{O}$\text{ is }$(k-u)\times\frac{q^u-1}{q-1}$ zero matrix, they have obtained 
 \begin{equation}\label{mac}
 G_{k,u}(q) = \left(\begin{matrix}
                   G_k(q) \setminus \left(\begin{matrix}
0\\
G_u(q)
\end{matrix}\right)
\end{matrix} \right)
 \end{equation}
 for $2 \leq u \leq k-1$  and $\left(\begin{matrix}
                   A \setminus B
\end{matrix} \right)$ is a matrix obtained from the matrix $A$ by removing the matrix $B.$  A code
generated by the matrix $G_{k,u}(q)$ is called {\it $\mathbb Z_{q}$-MacDonald} code. It is denoted
by $M_{k,u}(q).$ It is clear that the dimension of this code is $k.$ The Quaternary
MacDonald codes were discussed in \cite{cg03} and the MacDonald codes over finite field were
discussed in \cite{bhandari}.
 
In this correspondence, we concentrate on $\mathbb Z_{q}$-MacDonald Code. In Section $2$, we
determine the parameters of $\mathbb Z_q$-MacDonald code of dimension $k$ and in Section $3$, we obtain the weight distribution of $\mathbb Z_q$-MacDonald code of dimension $3$, for any positive integer $q\geq 2$.
In Section $4$, we find the weight distribution of $\mathbb Z_q$-Simplex code of dimension $3$, for any positive integer $q\geq 2.$ 

\section{Minimum Distance of $\mathbb Z_{q}$-MacDonald Code of dimension $k$}
In Equation 1.1, if we put $u = k-1,$ then a generator matrix of $k$-dimensional $\mathbb Z_{q}$-MacDonald code is
$$G_{k,k-1}(q) = \left[\begin{array}{c|c|c|c|c}
   1      &1 1 \cdots 1& 2 2\cdots 2 & \cdots & q-1q-1 \cdots q-1 \\\hline
               0      &            &             &        &\\
          \vdots & G_{k-1}(q)      &G_{k-1}(q)        &\cdots  &G_{k-1}(q)\\
               0      &            &             &        &\\
            \end{array}\right] $$            
where $G_{k-1}(q)$ is a generator matrix of $\mathbb {Z}_q$-Simplex code of dimension $k-1$. Then this matrix generates the code
\begin{equation*}
M_{k, k-1}(q)=\{(0cc \cdots c)+\alpha (1{\bf 12 \cdots q-1})\mid \alpha \in \mathbb {Z}_q, ~c\in S_{k-1}(q)\}  
\end{equation*}
 where ${\bf i}=ii \cdots i\in \mathbb Z_{q}^{n}$ and $n=\frac{q^{k-1}-1}{q-1}=n_{k-1}$.
The code generated by the Matrix $G_{k, k-1}(q)$ is a $[q^{k-1}, ~k, ~d(M_{k, k-1}(q))]$ $\mathbb {Z}_q$-linear code.\\ 
\noindent{\bf Case (i).} Let $\alpha =0.$ Then
\begin{equation}
\min\{wt(0cc \cdots c)\mid c \in S_{k-1}(q)\}=(q-1)d(S_{k-1}(q))=(q-1)\Big(\frac{q}{p}(p-1)n_{k-2}+1\Big) 
\end{equation}
where $p>1$ is the smallest divisor of $q$.\\
\noindent{\bf Case (ii).} Let $\alpha \neq0$. \\
\noindent{\bf Subcase (i).} Let $\alpha \in \mathbb {Z}_q$ with $(\alpha, q)=1.$ If $\alpha i=\alpha j,$ then $\alpha(i-j)=0.$ Since $\alpha$ is a unit, it implies $i=j.$ Therefore 
$\{\alpha {\bf 1}, ~\alpha {\bf 2}, ~\cdots, ~\alpha {\bf (q-1)}\}=\{{\bf 1}, ~{\bf 2}, \cdots, ~{\bf q-1}\}$.\\  
Consider
\begin{eqnarray*}
wt\big((0cc \cdots c)+\alpha (1{\bf 12 \cdots q-1})\big)&=&wt\big((0cc \cdots c)+(\alpha ~\alpha {\bf 1} ~\alpha {\bf 2} ~\cdots ~\alpha {\bf (q-1)})\big)\\
&=&wt\big((0cc \cdots c)+(\alpha ~{\bf 1 ~2 ~\cdots ~q-1})\big)\\
&=&1+\sum \limits_{i=1}^{q-1}wt(c+{\bf i})\\
wt\big((0cc \cdots c)+\alpha (1{\bf 12 \cdots q-1})\big)&=&1+\sum \limits_{i=1}^{q-1}wt(-c+{\bf i}) ~\text {since} ~S_{k-1}(q) ~\text {is} ~\mathbb {Z}_q\text{-linear}.
\end{eqnarray*}
Let $n(i)$ be the number of $i$ coordinates in $c \in S_{k-1}(q)$ where $i=0, ~1, ~2, ~\cdots, ~q-1.$ Then for $0\leq i\leq q-1$, $wt(-c+{\bf i})=n-n(i),$
where $n$ is the length of $S_{k-1}(q).$ Therefore,
\begin{eqnarray*}
wt\big((0cc \cdots c)+\alpha (1{\bf 12 \cdots q-1})\big)&=&1+\sum \limits_{i=1}^{q-1} (n-n(i))\\
&=&1+(q-1)n-\sum \limits_{i=1}^{q-1} n(i)\\
&=&1+(q-1)n-(n-n(0))\\
wt\big((0cc \cdots c)+\alpha (1{\bf 12 \cdots q-1})\big)&=&1+(q-2)n+n(0) ~\text {for all} ~c \in S_{k-1}(q).
\end{eqnarray*}
Therefore,
\begin{eqnarray*}
\min\Big\{wt((0cc \cdots c)+\alpha (1{\bf 12 \cdots q-1}))\mid c \in S_{k-1}(q), ~\alpha \in U(\mathbb {Z}_q)\Big\}
&=&1+(q-2)n+\min \limits_{c \in S_{k-1}(q)}\{n(0)\},
\end{eqnarray*}
where $U(\mathbb {Z}_q)$ is the set of all units in $\mathbb Z_{q}$. Thus, we have
\begin{equation}
\min\big\{wt\big((0cc \cdots c)+\alpha (1{\bf 12 \cdots q-1})\big)\mid(\alpha, ~q)=1\big\}=1+(q-2)n+\min \limits_{c \in S_{k-1}(q)}\{n(0)\}
\end{equation}
The largest weight codeword of $S_{k-1}(q)$ gives the minimum value of the above Equation 2.2.\\
\noindent{\bf Subcase (ii).} Let $(\alpha, q) \neq1$ and $o(\alpha)=d$. Then, $\{\alpha {1}, ~\alpha {2}, ~\cdots, ~\alpha (q-1)\}=\{\alpha {1}, ~\alpha {2}, ~\cdots, ~\alpha (d-1), ~0\}.$ Clearly, in 
$\{\alpha {1}, ~\alpha {2}, ~\cdots, ~\alpha (q-1)\}$, each non-zero $\alpha {i}$ appears $\frac{q}{d}$ times and zero appears $(\frac{q}{d}-1)$ times.\\
Consider
\begin{equation*}
wt\big((0cc \cdots c)+\alpha (1{\bf 12 \cdots q-1})\big)=wt\big((0cc \cdots c)+(\alpha ~\alpha {\bf 1} ~\alpha {\bf 2} ~\cdots ~\alpha ({\bf q-1}))\big)\\
\end{equation*}
\begin{eqnarray}
wt\big((0cc \cdots c)+\alpha (1{\bf 12 \cdots q-1})\big)&=&1+\frac{q}{d}\Big\{wt(\alpha{\bf 1}+c)+wt(\alpha{\bf 2}+c)+ \cdots +wt(\alpha({\bf d-1})+c)\Big\} \nonumber \\
&&+\big(\frac{q}{d}-1\big)wt(c)
\end{eqnarray}
If $c_{i}\notin <\alpha>$ for all ~$i$, then $wt(\alpha{i}+c)=n$. Therefore, Equation 2.3 becomes
\begin{equation}
wt\big((0cc \cdots c)+\alpha (1{\bf 12 \cdots q-1})\big)=1+\frac{q}{d}\big(d-1\big)n+\big(\frac{q}{d}-1\big)wt(c) 
\end{equation}
If $c_{i}\in <\alpha>$ for a few $r$ ~$c_i$'s, Equation 2.3 becomes
\begin{eqnarray}
wt\big((0cc \cdots c)+\alpha (1{\bf 12 \cdots q-1})\big)&=&1+\frac{q}{d}\Big[(d-1-r)n+(n-n(c_{1}))+(n-n(c_{2}))+\cdots + \nonumber\\
&&(n-n(c_{r}))\Big]+\big(\frac{q}{d}-1\big)wt(c)\nonumber\\\
wt\big((0cc \cdots c)+\alpha (1{\bf 12 \cdots q-1})\big)&=&1+\frac{q}{d}\Big[(d-1)n-\sum \limits_{i=1}^{r}n(c_{i})\Big]+\big(\frac{q}{d}-1\big)wt(c)
\end{eqnarray}
If there is a $c \in S_{k-1}(q)$ such that $c_{i}\in <\alpha>$ for all ~$i$, then $\sum \limits_{i=1}^{d}n(c_{i})=n$. Therefore, Equation 2.3 becomes
\begin{eqnarray*}
wt\big((0cc \cdots c)+\alpha (1{\bf 12 \cdots q-1})\big)&=&1+\frac{q}{d}\Big[(n-n(c_{1}))+(n-n(c_{2}))+\cdots +\nonumber\\
&&(n-n(c_{d-1}))\Big]+\big(\frac{q}{d}-1\big)wt(c)\nonumber\\
&=&1+\frac{q}{d}\Big[(n-n(\alpha))+(n-n(2\alpha))+\cdots +\nonumber\\
&&(n-n((d-1)\alpha))\Big]+\big(\frac{q}{d}-1\big)wt(c)\nonumber\\
wt\big((0cc \cdots c)+\alpha (1{\bf 12 \cdots q-1})\big)&=&1+\frac{q}{d}\Big[(d-1)n-\sum \limits_{i=1}^{d-1}n(i\alpha)\Big]+\big(\frac{q}{d}-1\big)wt(c)
\end{eqnarray*}
Since $n=\sum \limits_{i=1}^{d-1}n(c_{i})+n(0)$, we get
\begin{eqnarray}
wt\big((0cc \cdots c)+\alpha (1{\bf 12 \cdots q-1})\big)&=&1+\frac{q}{d}\Big[(d-1)n-(n-n(0))\Big]+\Big(\frac{q}{d}-1\Big)wt(c)\nonumber\\
&=&1+\frac{q}{d}\Big[(d-2)n+n(0)\Big]+\Big(\frac{q}{d}-1\Big)wt(c)\nonumber\\
&=&1+\frac{q}{d}\Big[(d-2)n+n(0)\Big]+\Big(\frac{q}{d}-1\Big)\Big(n-n(0)\Big), \nonumber\\
&&~\text {where} ~c_{i}\in <\alpha>\nonumber\\
&=&1+\frac{q}{d}\big(d-2\big)n+\frac{q}{d}n(0)+\frac{q}{d}n-n-\frac{q}{d}n(0)+n(0)\nonumber\\
wt\big((0cc \cdots c)+\alpha (1{\bf 12 \cdots q-1})\big)&=&1+(q-1)n-\frac{q}{d}n+n(0) 
\end{eqnarray}
%
From Equations 2.4, 2.5 and 2.6, we have
\begin{equation*}
1+\frac{q}{d}\Big[(d-2)n+n(0)\Big]+\Big(\frac{q}{d}-1\Big)wt(c)\leq wt\Big[(0cc \cdots c)+\alpha (1{\bf 12 \cdots q-1})\Big]\leq 1+\frac{q}{d}\Big[(d-1)n\Big]+\Big(\frac{q}{d}-1\Big)wt(c) 
\end{equation*}
for all $c \in S_{k-1}(q)$ ~\text{and} $(\alpha, ~q)\neq1$.

If there exists $c \in S_{k-1}(q)$ such that $c_{i}\in <\alpha>$ and $d$ is smaller, then the Equation 2.6 gives the smaller value. That is,
\begin{equation}
\min \limits_{c \in S_{k-1}(q)}\Big\{wt\big((0cc \cdots c)+\alpha (1{\bf 12 \cdots q-1})\big)\mid (\alpha, ~q)\neq1\Big\}=1+\big(q-\frac{q}{d}-1\big)n+\Big\{\min \limits_{c\in S_{k-1}(q)} n(0)\Big\} 
\end{equation}
where $n(0)$ is the number of zeros in $c$ such that $c_{i}\in <\alpha>$ and $\alpha$ must be smaller order element. Therefore,
\begin{eqnarray*}
\min \Big\{wt\big((0cc \cdots c)+\alpha (1{\bf 12 \cdots q-1})\big) \mid \alpha \in \mathbb {Z}_q, ~c \in S_{k-1}(q)\Big\}
&=&\min\Big\{(q-1)d(S_{k-1}(q)), ~1+(q-2)n+\\
&&\min \limits_{c \in S_{k-1}(q)}\{n(0)\}, ~1+\big(q-\frac{q}{d}-1\big)n+\\
&&\min \limits_{c \in S_{k-1}(q)}n(0)\Big\}\\ 
&=&\min \Big\{(q-1)d(S_{k-1}(q)), \\
&&~1+\big(q-\frac{q}{d}-1\big)n+\min \limits_{c \in S_{k-1}(q)}n(0)\Big\} 
\end{eqnarray*}
Let $\alpha$ be a least order non-zero element in $\mathbb {Z}_q$. Since $011 \cdots 1 \in S_{2}(q)$, it implies that $c=0\alpha\alpha \cdots \alpha \in S_{2}(q).$
Therefore, for $k=3$, the above Equation 2.7 becomes
\begin{equation*}
\min \limits_{c \in S_{2}(q)}\{wt(0cc \cdots c)+\alpha (1{\bf 12 \cdots q-1)}\}=1+(q-\frac{q}{d}-1)n_{2}+n(0) 
\end{equation*}
Since $n(0)=1,$ it implies that,
\begin{equation*}
\min \limits_{c \in S_{2}(q)}\big\{wt\big((0cc \cdots c)+\alpha (1{\bf 12 \cdots q-1})\big)\big\}=2+\big(q-\frac{q}{d}-1\big)n_{2} 
\end{equation*}
For $k=3$, Equation 2.1 gives $\min \limits_{c\in S_{2}(q)}\{wt(0cc\cdots c)\}=(q-1)d(S_{2}(q))$ and Equation 2.2 gives 
$\min \limits_{c\in S_{2}(q)}\{wt(0cc\cdots c)+\alpha (1{\bf 12 \cdots q-1)}\}=1+(q-2)n_{2}+1=2+(q-2)n_{2}$.
Therefore, the minimum distance of $M_{3,2}(q)$ is 
\begin{equation*}
d(M_{3,2}(q))=\min \big\{(q-1)d(S_{2}(q)), ~2+\big(q-\frac{q}{d}-1\big)n_{2}\big\}. 
\end{equation*}
Since $d(S_{2}(q))=\frac{q}{p}(p-1)+1$, it follows that $d(M_{3,2}(q))=2+\big(q-\frac{q}{d}-1\big)n_{2}$.\\
For $k=4$, in $S_{3}(q)$, the codeword $c=0\alpha\alpha \cdots \alpha \in S_{2}(q)$ is repeated $q$ times and $c^{'}=c0c\cdots c$ is a codeword in $S_{3}(q)$ 
which gives the minimum number of zeros, and all coordinates of $c^{'}$ are in $<\alpha>$. The number of zeros in $c^{'}$ is $q+1$. Hence, Equation 2.7 becomes
\begin{equation*}
\min \limits_{c \in S_{3}(q)}\big\{wt\big((0cc \cdots c)+\alpha (1{\bf 12 \cdots q-1})\big)\big\}=1+\big(q-\frac{q}{d}-1\big)n_{3}+(q+1). 
\end{equation*}
For $k=5$, the codeword $c^{'} \in S_{3}(q)$ is repeated $q$ times in $S_{4}(q)$ and hence $c^{''}=c^{'}0c^{'}\cdots c^{'}$ is codeword in $S_{4}(q)$ 
which gives the minimum number of zeros, and its coordinates are in $<\alpha>$. The number of zeros in $c^{''}$ is $[q(q+1)]+1$. Hence, Equation 2.7 becomes
\begin{equation*}
\min \limits_{c \in S_{4}(q)}\big\{wt\big((0cc \cdots c)+\alpha (1{\bf 12 \cdots q-1})\big)\big\}=1+\big(q-\frac{q}{d}-1\big)n_{4}+\big[q(q+1)+1\big]. 
\end{equation*}
In general, for any $k$, in $S_{k-1}(q),$ there is a codeword $c \in S_{k-2}(q)$ the coordinates of which are in $<\alpha>$ with minimum number of zeros $\frac{q^{k-3}-1}{q-1}$ and hence $c_{1}=c0c\cdots c$ is a codeword in $S_{k-1}(q)$ 
which gives the minimum number of zeros, and its coordinates are in $<\alpha>$. The number of zeros in $c_{1}$ is $\frac{q^{k-2}-1}{q-1}$. Hence, Equation 2.7 becomes
\begin{equation*}
\min \limits_{c_{1} \in S_{k-1}(q)}\big\{wt\big((0c_{1}c_{1} \cdots c_{1})+\alpha (1{\bf 12 \cdots q-1})\big)\big\}=1+\big(q-\frac{q}{d}-1\big)n_{k-1}+\frac{q^{k-2}-1}{q-1}. 
\end{equation*} 
Therefore, 
\begin{equation*}
d(M_{k, k-1}(q))=1+\big(q-\frac{q}{d}-1\big)n_{k-1}+\frac{q^{k-2}-1}{q-1}. 
\end{equation*} 
Thus, we have\\
\begin{theorem}
 The $\mathbb {Z}_q$-MacDonald code $M_{k, k-1}(q)$ is a $\big[q^{k-1}, ~k, ~1+\big(q-\frac{q}{d}-1\big)\big(\frac{q^{k-1}-1}{q-1}\big)+\frac{q^{k-2}-1}{q-1}\big]$ $\mathbb {Z}_q$-linear code
 where $d>1$ is the smallest divisor of $q$.
\end{theorem}

\section{Weight Distribution of $\mathbb {Z}_q$-MacDonald Code of Dimension 3}

Let
$$G_{3, 2}(q)= \left[\begin{array}{c|l|l|c|l}
1& 1 ~1 ~1 ~1 ~\cdots \ \ \ 1& 2 ~2 ~2 ~2 ~\cdots \ \ \ 2 & ~\cdots & q-1 q-1 ~q-1 ~q-1 ~~\cdots ~~~~~q-1 \\\hline
0& 0 ~1 ~1 ~2 ~\cdots ~q-1& 0 ~1 ~1 ~2 ~\cdots ~q-1& ~\cdots & 0 ~~~~~~1 ~~~~~~1 ~~~~~~2 ~~~~~~\cdots ~~~~~~q-1\\
0& 1 ~0 ~1 ~1 ~\cdots \ \ \ 1& 1 ~0 ~1 ~1 ~\cdots \ \ \ 1  & ~\cdots & 1 ~~~~~~0 ~~~~~~1 ~~~~~~1 ~~~~~~\cdots ~~~~~~\ \ 1\end{array}\right] $$
Then by Theorem 2.1, this matrix generates $\big[q^{2}, ~3, ~2+\big(q-\frac{q}{d}-1\big)\big(q+1\big)\big]$ $\mathbb {Z}_q$-linear code where $d>1$ is the smallest divisor of $q$. It is 
$M_{3, 2}(q)=\{(0cc \cdots c)+\alpha(1{\bf 12 \cdots q-1}) \mid \alpha \in \mathbb {Z}_q\}$.  
In \cite{r7}, they have given the weight distribution of $2$-dimensional $\mathbb Z_{q}$-Simplex code as 
\begin{theorem}\cite{r7}
 For any integer $q\geq 2,$ the weight distribution of $\mathbb Z_{q}$-Simplex code of dimension $2$ is
\[
\begin{array}{ccl}
A_{0}&=&1 \\
A_{q}&=&q\phi(q)+q-1\\
A_{q-\frac{q}{d}+1}&=& d\phi (d),
\text{ for } d|q \text{ and } d \neq1, d \neq q \\
A_{q +1}&=& q(q-1)-\sum \limits_{{d|q}, {d \neq 1}}d\phi (d) .
\end{array}
\]
where $d>1$ is the smallest divisor of $q$.
\end{theorem}
Note that there is only one codeword in $S_{2}(q)$ such that $n(0)=n$, $d\phi (d)$ codewords in $S_{2}(q)$ such that $n(0)=\frac{q}{d}$, ~for all ~$d|q$, $d\neq1$ and $~d\neq q$, 
$q\phi(q)+q-1$ codewords in $S_{2}(q)$ such that $n(0)=1$ and $q(q-1)-\sum \limits_{{d|q}, {d \neq 1}}d\phi (d)$ codewords in $S_{2}(q)$ such that $n(0)=0$.\\ 
Now, we consider the code $M_{3, 2}(q)$.\\ 
\noindent{\bf Case (i).} If $\alpha=0$, then
\begin{eqnarray}
wt\big((0cc \cdots c)+\alpha (1{\bf 12 \cdots q-1})\big)&=&wt(0cc \cdots c)\nonumber\\
&=&(q-1)wt(c), ~\text {for all} ~c \in S_{2}(q)
\end{eqnarray}
By Theorem 3.1, we get the following weights:
\begin{equation}
\begin{cases}
\text {Number of zero weight codeword is} ~1.\\
\text {Number of} ~(q-1)(q-\frac{q}{d}+1) ~\text {weight codeword is} ~d\phi (d) ~\text {where} ~d|q, ~d\neq1 ~\text {and} ~d\neq q.\\
\text {Number of} ~(q-1)q ~\text {weight codeword is} ~q\phi(q)+q-1.\\
\text {Number of} ~(q-1)(q+1)=q^{2}-1 ~\text {weight codeword is} ~q(q-1)-\sum \limits_{{d|q}, {d \neq 1}}d\phi (d). 
\end{cases}
\end{equation}
\noindent{\bf Case (ii).} If $(\alpha, q)=1,$ then by Equation 2.2,
\begin{equation}
wt\big((0cc \cdots c)+\alpha (1{\bf 12 \cdots q-1})\big)=1+(q-2)n+n(0),
\end{equation}
for all $c \in S_{2}(q)$, and $n(0)$ is the number of zeros in $c$.
By Theorem 3.1, in $M_{3, 2}(q)$,
\begin{equation}
\begin{cases}
\text {Number of}  ~1+(q-1)n ~\text {weight codeword is} ~1.\phi (q).\\
 \text {Number of} ~1+(q-2)n+\frac{q}{d} ~\text {weight codeword is} ~(d\phi (d)).\phi (q), ~\text {for all} ~d|q, d\neq1 ~\text {and} ~d\neq q.\\
 \text {Number of} ~2+(q-2)n ~\text {weight codeword is} ~(q\phi(q)+q-1).\phi (q).\\
 \text {Number of}  ~1+(q-2)n ~\text {weight codeword is} ~(q(q-1)-\sum \limits_{{d|q}, {d \neq 1}}d\phi (d)).\phi (q).
 \end{cases}
\end{equation}
\noindent{\bf Case (iii).} If $\alpha$ is not relatively prime to $q$, then by Equation 2.6, we have
\begin{equation}
wt((0cc \cdots c)+\alpha (1{\bf 12 \cdots q-1}))=1+(q-1)n-\frac{q}{d}\sum \limits_{i=0}^{d-1}n({\bf i}\alpha)+n(0) 
\end{equation}
where $n({\bf i}\alpha)$ is the number of ${\bf i}\alpha$'s in $c$.\\
If we know the details of coordinates in $c$, we can get the remaining weights of $M_{3, 2}(q).$
\begin{example}
For $q=4, k=3,$ the Matrix
$$G_{3, 2}(4)= \left[\begin{array}{c|c|c|c} 1&1 1 1 1 1& 2 2 2 2 2& 3 3 3 3 3 \\\hline
0&0 1 1 2 3& 0 1 1 2 3& 0 1 1 2 3\\
0&1 0 1 1 1& 1 0 1 1 1& 1 0 1 1 1\end{array}\right] $$
\end{example}
generates the code 
\begin{equation*}
M_{3,2}(4)=\{(0ccc)+\alpha(1{\bf 123})\mid \alpha \in \mathbb {Z}_4\}  
\end{equation*}
By Theorem 3.1, the weight distribution of $S_{2}(4)$ is
\begin{center}
$A_{0}=1, ~A_{4}=11, ~A_{3}=2, ~A_{5}=2$ 
\end{center}
and hence the $n(0)$s are such that 5, ~1, ~2 and ~0 respectively.\\
\noindent{\bf Case (i).} If $\alpha=0,$ then using Equation 3.1, we have 
\begin{center}
$wt((0ccc)+\alpha(1{\bf 123}))=3wt(c)$, 
\end{center}
for all $c \in S_{2}(4).$ Therefore, by Equation 3.2, there is only one codeword of weight zero, 2 codewords of weight 9, 
11 codewords of weight 12 and 2 codewords of weight 15.\\
\noindent{\bf Case (ii).} If $(\alpha, 4)=1,$ then $\alpha \in \{1, 3\}$, by using Equation 3.3, we have,
\begin{center}
$wt((0ccc)+\alpha(1{\bf 123}))=1+(2)(5)+n(0)=11+n(0)$
\end{center}
By Equation 3.4, there are 2 codewords of weight 16, 4 codewords of weight 13, 
22 codewords of weight 12 and 4 codewords of weight 11.\\
\noindent{\bf Case (iii).} If $\alpha$ is not relatively prime to 4, then $\alpha \in \{2\}$ and by Equation 3.5, we get,
\begin{eqnarray*}
wt((0ccc)+2(1{\bf 123}))&=&1+(4-1)(5)-\frac{4}{2}\sum \limits_{i=0}^{1}n(i2)+n(0)\\
&=&1+15-2[n(0)+n(2)]+n(0)\\
wt((0ccc)+2(1{\bf 123}))&=&16-n(0)-2n(2).
\end{eqnarray*}
Using the coordinates of $c \in S_{2}(q)$, there is only one codeword of weight 11, only one codeword of weight 7, 2 codewords of weight 8, 
8 codewords of weight 13,  2 codewords of weight 14 and 2 codewords of weight 15.\\
By combining cases (i), (ii) and (iii), we have
\begin{theorem}
The weight distribution of $\mathbb {Z}_4$-MacDonald code $M_{3,2}(4)$ is
\begin{center}
$A_{0}=1, ~A_{7}=1, ~A_{8}=2, ~A_{9}=2, ~A_{11}=5, ~A_{12}=33, ~A_{13}=12$, $A_{14}=2, ~A_{15}=4, ~A_{16}=2.$
\end{center}
\end{theorem}
\section{Weight Distribution of $\mathbb{Z}_q$-Simplex Code of dimension 3, for any $q\geq 2$.}
Let
$$G_{3}(q)= \left[\begin{array}{l|c|l|l|c|l}
0 ~0 ~0  ~\cdots ~~~~0&1& 1 ~1 ~1 ~\cdots ~~~~1& 2 ~2 ~2 ~\cdots ~~~~2 & ~\cdots & q-1 ~q-1 ~q-1  ~\cdots ~~~~~q-1 \\\hline
0 ~1 ~1 ~\cdots ~q-1&0& 0 ~1 ~1 ~\cdots ~q-1& 0 ~1 ~1 ~\cdots ~q-1& ~\cdots & 0 ~~~~~~1 ~~~~~~1 ~~~~~~\cdots ~~~~~~q-1\\
1 ~0 ~1 ~\cdots ~~~~1&0& 1 ~0 ~1 ~\cdots ~~~~1& 1 ~0 ~1 ~\cdots ~~~~1  & ~\cdots & 1 ~~~~~~0 ~~~~~~1 ~~~~~~\cdots ~~~~~~~~~1\end{array}\right] $$
Then this matrix generates the code $S_{3}(q)=\{(c0c \cdots c)+\alpha({\bf 0}1{\bf 12 \cdots q-1}) \mid \alpha \in \mathbb {Z}_q\}$.\\  
In \cite{r6}, we have given the parameters of $S_{k}(q)$, and the weight distribution of $S_{2}(q)$ is given by Theorem 3.1.\\
\noindent{\bf Case (i).} Let $\alpha=0$. Then,
\begin{eqnarray}
wt((c0cc \cdots c)+\alpha ({\bf 0}1{\bf 12 \cdots q-1}))&=&wt(c0cc \cdots c)\nonumber\\
&=&(q)wt(c), ~\text {for all} ~c \in S_{2}(q)
\end{eqnarray}
In this way, we get the following weights.
\begin{enumerate}
\item Number of zero weight codeword is $1$.
\item Number of $q\big(q-\frac{q}{d}+1\big)$ weight codeword is $d\phi (d)$, where $d|q,$ $d\neq1$ and $d\neq q$.
\item Number of $qq=q^{2}$ weight codeword is $q\phi(q)+q-1$.
\item Number of $q(q+1)$ weight codeword is $q(q-1)-\sum \limits_{{d|q}, {d \neq 1}}d\phi (d)$. 
\end{enumerate}
\noindent{\bf Case (ii).} Let $(\alpha, q)=1.$ Since $\{\alpha.1, ~\alpha.2, ~\cdots, ~\alpha.(q-1)\}=\{1, ~2, ~\cdots, ~q-1\},$
\begin{eqnarray}
wt\big((c0cc \cdots c)+\alpha ({\bf 0}1{\bf 12 \cdots q-1})\big)&=&wt\big((c0cc \cdots c)+({\bf 0}\alpha{\bf 12 \cdots q-1}\big) \nonumber\\
&=&1+\sum \limits_{i=0}^{q-1}wt\big(c+{\bf i}\big)\nonumber\\
&=&1+\sum \limits_{i=0}^{q-1}wt\big(-c+{\bf i}\big)\nonumber\\
&=&1+\sum \limits_{i=0}^{q-1}\big[n-n({\bf i})\big]\nonumber\\
&=&1+qn-n\nonumber\\
wt\big((c0cc \cdots c)+\alpha ({\bf 0}1{\bf 12 \cdots q-1})\big)&=&1+(q-1)n ~\text {for all} ~c \in S_{2}(q).
\end{eqnarray}
From the above, the number of $1+(q-1)n$ weight codeword is $\#\big(S_{2}(q)\big)=q^{2}$ for all $c \in S_{2}(q)$.

Since there are $\phi (q) ~\alpha$'s such that $(\alpha, q)=1$, it implies that the number of $1+(q-1)n$ weight codeword is $\phi (q).q^{2}$ and hence
\begin{equation}
A_{1+(q-1)n}=\phi (q).q^{2}  
\end{equation}
\noindent{\bf Case (iii).} If $\alpha$ is not relatively prime to $q$, then
\begin{eqnarray*}
wt\big((c0cc \cdots c)+\alpha ({\bf 0}1{\bf 12 \cdots q-1})\big)&=&1+wt(c+\alpha{\bf 0})+wt(c+\alpha{\bf 1})+ \cdots +wt(c+\alpha({\bf q-1}))\\
&=&1+wt(0\alpha-c)+wt(1\alpha-c)+ \cdots +wt((d-1)\alpha-c)+ \cdots \\
&=&1+\frac{q}{d}\sum \limits_{i=0}^{d-1}wt\big({\bf i}\alpha-c\big)\\
&=&1+\frac{q}{d}\sum \limits_{i=0}^{d-1}\big[n-n({\bf i}\alpha)\big]\\
&=&1+\frac{q}{d}\big[dn\big]-\frac{q}{d}\sum \limits_{i=0}^{d-1}n({\bf i}\alpha)\\
&=&1+qn-\frac{q}{d}\sum \limits_{i=0}^{d-1}n({\bf i}\alpha)
\end{eqnarray*}
Therefore,
\begin{equation}
wt\big((c0cc \cdots c)+\alpha ({\bf 0}1{\bf 12 \cdots q-1})\big)=1+qn-\frac{q}{d}\sum \limits_{i=0}^{d-1}n({\bf i}\alpha), 
\end{equation}
where $n({\bf i}\alpha)$ is the number of ${\bf i}\alpha$'s in $c$. 
If we know the details of coordinates in $c$, we can get the remaining weights of $S_{3}(q).$
\begin{example}
For $q=4, k=3,$ the Matrix
$$G_{3}(4)= \left[\begin{array}{c|c|c|c|c} 0 0 0 0 0&1&1 1 1 1 1& 2 2 2 2 2& 3 3 3 3 3 \\\hline
0 1 1 2 3&0&0 1 1 2 3& 0 1 1 2 3& 0 1 1 2 3\\
1 0 1 1 1&0&1 0 1 1 1& 1 0 1 1 1& 1 0 1 1 1\end{array}\right] $$
\end{example}
generates the code 
\begin{equation*}
S_{3}(4)=\{(c0ccc)+\alpha({\bf 0}1{\bf 123})\mid \alpha \in \mathbb {Z}_4, ~c \in S_{2}(4)\} ~\text {where} ~{\bf i}=ii \cdots i \in \mathbb Z_{q}^{n}  
\end{equation*}
\noindent{\bf Case (i).} Let $\alpha=0$. Then, using Equation 4.1, we have 
\begin{center}
$wt\big((c0ccc)+\alpha({\bf 0}1{\bf 123})\big)=wt(c0ccc)=4wt(c)$, 
\end{center}
for all $c \in S_{2}(4).$ Therefore, by using the weight distribution of $S_{2}(4),$ there is only one codeword of weight zero, 11 codewords of weight 16, 
2 codewords of weight 12 and 2 codewords of weight 20.\\
\noindent{\bf Case (ii).} If $(\alpha, 4)=1,$ then $\alpha \in \{1, 3\}$ and by Equation 4.2, we get
\begin{center}
$wt\big((c0ccc)+\alpha({\bf 0}1{\bf 123})\big)=1+3n$
\end{center}
Then, by Equation 4.3, the number of $1+3n=16$ weight codeword is $\phi (4).4^{2}=32$. That is, there are 32 codewords of weight 16.\\
\noindent{\bf Case (iii).} If $\alpha$ is not relatively prime to 4, then $\alpha \in \{2\}$ and by Equation 4.4, we get 
\begin{eqnarray*}
wt\big((c0ccc)+2({\bf 0}1{\bf 123})\big)&=&1+(4)(5)-\frac{4}{2}\sum \limits_{i=0}^{1}n(i2)\\
&=&21-2[n(0)+n(2)].
\end{eqnarray*}
Using the coordinates of $c \in S_{2}(4),$ we get, there are 4 codewords of weight 11, 8 codewords of weight 17 and 4 codewords of weight 19.
Therefore, by combining cases (i), (ii) and (iii), we have
\begin{theorem}
The weight distribution of $\mathbb {Z}_4$-Simplex code of dimension 3 is 
\begin{center}
$A_{0}=1, ~A_{11}=4, ~A_{12}=2, ~A_{16}=43, ~A_{17}=8, ~A_{19}=4, ~A_{20}=2$.
\end{center}
\end{theorem}
\section*{Acknowledgments}

For the first author, this research has been supported by the Women Scientists Scheme-A, Department of Science and Technology, Government of India grant DST: SR/WOS-A/MS-12/2012(G), dated: November 1st, 2012.


\begin{thebibliography}{0}
\bibitem{r1}
A.A. de Andrade and R. Palazzo,
{\it Linear Codes over Finite Rings, }
{TEMA Tend. Mat. Apl. Comput.,} {6(2)}, (2005), 207--217.


\bibitem{bhandari} Bhandari M. C. and Durairajan C.,  {\it A Note on covering radius of MacDonald Codes,} Proceeding of the
International Conference on Information Technology: Computers and Communication(ITCC-2003).

\bibitem{bgl99}  M. C. Bhandari, M. K. Gupta and A. K. Lal,
{ On $\mathbb Z_4$ Simplex codes and their gray images, } 
{\it Applied Algebra, Algebraic Algorithms and Error-Correcting Codes, AAECC-13, 
Lecture Notes in Computer Science, }
{1719} (1999) pp. 170--180.

\bibitem{byi}
Bahattin Yildiz,
{\it Weights modulo $p^{e}$ of linear codes over rings,} {Designs, Codes and Cryptography, } {43}, (2007), 147--165.


%

\bibitem{cpcd} P. Chella Pandian and C. Durairajan, {On  $\mathbb Z_{q}$-linear and $\mathbb Z_{q}$-Simplex codes and its re-
lated parameters for $q$ is a prime power}, { {Journal of Discrete Mathematical Sciences and Cryptography},} {Vol. 18 (2015), No. 1 and 2,} pp. 81--94.

\bibitem{cg03} Colbourn C. J. and Gupta M. K., {\it On quaternary MacDonald codes,} Proc.
Information Technology  Coding and Computing (ITCC), April, 2003, pp. 212--215.        

\bibitem{cs93}  J. H. Conway and  N. J. A. Sloane,
{ Self-dual codes over the integers modulo 4, }
{\it Journal of Combinatorial Theory Series A, }
{62} (1993) pp. 30--45.

%

\bibitem{dk10}  S.T. Dougherty and J.L. Kim, { Construction of self-dual codes over chain rings, }
{ \it Int. J. Inf. Coding Theory, }
{1}(2)  (2010) pp. 171--190.

\bibitem{r12}
S. T. Dougherty, T. Aaron Gulliver, Young Ho Park and John N.C. Wong,
{\it Optimal Linear codes over $\mathbb Z_{m}$, }
{J. Korean Math. Soc., } {44(5)}, (2007), 1139-1162.

\bibitem{r13}
S. T. Dougherty, Manish K. Gupta and Keisuke Shiromoto,
{\it On generalised weights for codes over $\mathbb Z_{k}$, }
{Australasian Journal of Combinatorics}, {31}, (2005), 231-248.

\bibitem{r6}
C. Durairajan, J. Mahalakshmi and P. Chella Pandian,
{\it On the $\mathbb Z_{q}$-Simplex Codes and its Weight Distribution for Dimension $2$, }Communicated.

\bibitem{r7}
C. Durairajan and J. Mahalakshmi ,
{\it On Codes over Integers Modulo $q$, } {\it Advances and Applications in Discrete mathematics,} Accepted.

\bibitem{eby}
Eimear Byrne,
{\it On the Weight Distribution of Codes over Finite Rings, }
{ arXiv.org $>$ math $>$ arXiv:1101.1505v1, } (2011).


\bibitem{r8}
Eugene Spiegel,
{\it Codes over $\mathbb Z_{m}$,}
{ Information and Control, } {35,} (1977), 48-51.

\bibitem{r9}
M. K. Gupta and C. Durairajan, {\it On the Covering Radius of some Modular Codes, }
{Advances in Mathematics of Communications,} {8(2)}, (2014), 129-137.

%



\bibitem{vsr96}
  V. V. Vazirani, H. Saran and  B. Sundar Rajan, 
{ An efficient algorithm for constructing minimal trellises for codes
 over finite abelian groups,} 
 {\it IEEE Trans. Inform. Theory}, Vol. 42, No. 6 (1996) pp. 1839--1854.

\end{thebibliography}
\end{document}